# Bound states in the continuum of infinite quality factor in finite unit cells

Huawei Liang[1], Yuanzhi Liu[1], Yu-Jia Zeng[1], Yangjian Cai[2], Tingyin Ning[2,*]

[1]Key Laboratory of Optoelectronic Devices and Systems of Ministry of Education and Guangdong Province, Shenzhen University, Shenzhen 518060, People's Republic of China

[2]Shandong Provincial Engineering and Technical Center of Light Manipulations & Shandong Provincial Key Laboratory of Optics and Photonic Device, School of Physics and Electronics, Shandong Normal University, Jinan 250358, China

**Abstract:**

A theory based on the superposition principle is developed to uncover the basic physics of the wave behavior in a finite grating of $N$ unit cells. The theory reveals that bound states in the continuum (BICs) of infinite quality factor (Q-factor) can be supported by such grating when the perfect reflection is introduced at its boundaries. If geometrical perturbations are introduced in the structure, the dark BICs transit to bright quasi-BICs of finite Q-factor, whose spectral behaviors are nearly the same as that of quasi-BICs supported by infinite gratings. When the boundaries are replaced with metallic mirrors of high reflectivity, the Q-factor of the resonant mode is reduced to be finite; however, it can be much larger than that in the corresponding nanostructure of open boundaries and can be tuned in a large range by varying the number of unit cells or boundary conditions.

Bound states in the continuum (BICs), with an infinite lifetime or quality factor (Q-factor), were first proposed in quantum mechanics as peculiar solutions of the Schrödinger equation with complex artificial potentials [1,2]. The concept of BICs was then extended to various physical systems, including acoustics [3-5], hydrodynamics [6-8], and photonics [9,10]. Especially, BICs in photonic systems have attracted substantial attentions as an emerging platform for advanced photonic applications, including the low threshold lasing [11-13], ultrasensitive biosensing [14-16], ultralow-level optical switching [17,18], highly efficient nonlinear optical response [19-22], efficient photon-pair generation [23,24], and chiral manipulation of photons [25], etc.

BICs of infinite Q-factor can be realized in extended structures which are uniform or periodic in one or more directions (in plane) [26]. In such systems, the translational symmetry conserves the in-plane wave vector $k_\parallel$, and thus a BIC state without radiation in the out-of-plane direction can exist inside the continuous spectrum of modes at the same $k_\parallel$. However, the periodic nanostructures of infinite unit cells can only be obtained theoretically. In practical applications, only finitely periodic nanostructures were fabricated [11,12,15,20]. Even if the loss in materials, structural defects, the effect of substrate, etc., can all be neglected, one basic physical question remains: can BICs of infinite Q-factor survive in the array of finite unit cells?

In previous studies, quasi-BICs or resonance modes with finite Q-factor in arrays of finite unit cells have been discussed [27-32]. The dependence of Q-factor on the number of unit cells $N$ follows a quadratic relation $Q \sim N^2$ for the symmetry-protected quasi-BICs [27-29], and a triple relation $Q \sim N^3$ for the accidental BICs [30,31]. However, BICs of infinite Q-factor have not been found in the finite arrays due to the open boundaries.

In this Letter, we establish a theory based on the superposition principle to describe the wave behavior in a finite grating of $N$ unit cells, which suggests that it can support BICs of infinite Q-factor by introducing the perfect reflection at the boundaries. The generation mechanism of the BICs is the combination of the symmetry protection and constructive interference of infinite waves caused by the perfect reflection, which mimics the interference process of the BICs in infinite unit cells. The dark BICs transit to bright quasi-BICs of high Q-factor when geometrical perturbations are introduced in the grating. Two branches of standing wave modes can be supported by the structure, whose spectral behaviors are nearly the same as that of quasi-BICs in infinitely periodic gratings. If the boundaries are replaced with metallic mirrors of high reflectivity, the Q-factor of the resonant mode becomes finite with or without the geometrical perturbations, which can be tuned in a large range by varying the number of unit cells or boundary conditions. The proposed theory provides in-depth understanding of BICs in finite unit cells, which have significant potentials in nano-photonic devices.

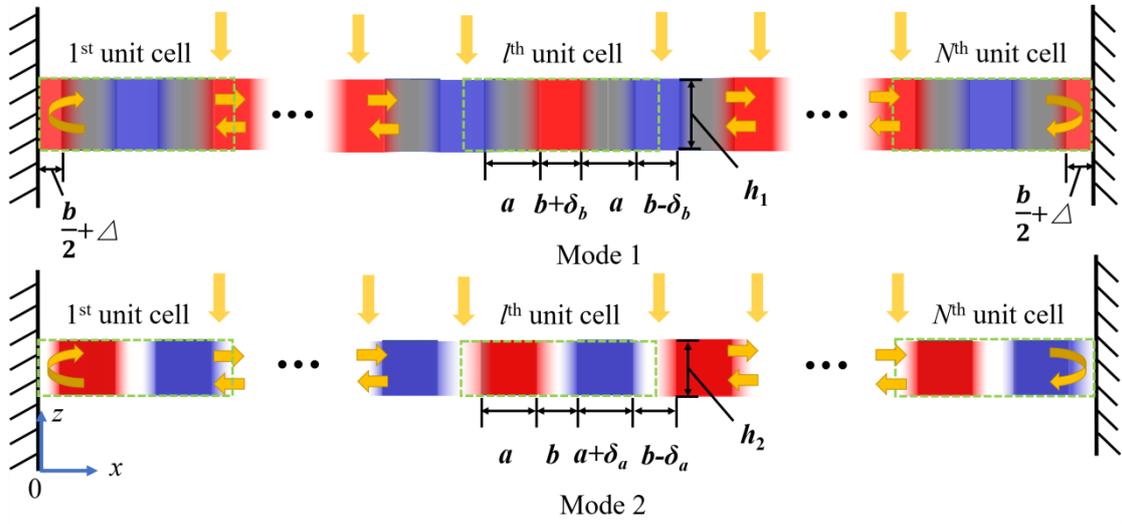

Fig. 1. Schematic diagram of 1D finite gratings of $N$ unit cells. Two types of geometrical perturbations are introduced to obtain two different quasi-BICs. Structure 1: the widths of two dielectric (Si) nanorods in a unit cell, $a$, remain unchanged, while the widths of air gaps become $b+\delta_b$ and $b-\delta_b$ by introducing a position shift, $\delta_b$. Structure 2: the widths of one dielectric (Si) nanorod and one air gap remain unchanged, while the widths of the other dielectric (Si) nanorod and air gap become $a+\delta_a$, and $b-\delta_a$, respectively, by introducing a perturbation, $\delta_a$. The width of air gaps between the outermost nanorods and the boundaries is $b/2 + \Delta$ with $\Delta = 0$ unless otherwise specified. The heights of dielectric (Si) nanorods $h_1 = 400$ nm and $h_2 = 120$ nm for attaining modes 1 and 2, respectively.

The configuration of one dimensional (1D) finite gratings of $N$ unit cells with boundaries at both ends is shown in Fig. 1. The periods of the gratings with and without the perturbation are both $d = 2a + 2b$, where $a$ and $b$ are the initial widths of the dielectric nanorods and air gaps without perturbations. $\delta_b$ and $\delta_a$ are geometrical perturbations in structures 1 and 2, respectively. The width of air gaps between the outermost nanorods and the boundaries is $b/2 + \Delta$ with $\Delta = 0$ unless otherwise specified. In the grating, two guided waves with propagation constants of $-\beta_m$ and $\beta_m$ propagate towards the left and right, respectively, where $m$ is the order number of the guided modes. Their initial amplitudes and phases are assumed $A_0$ and 0 in all unit cells, respectively. According to the coupled-mode theory, when the guided wave from the $l^{th}$ unit cell arrives at the right boundary of the $N^{th}$ unit cell, its electric field can be written as [33]

$$E_l = A_0 \frac{s'}{s' \cos\left[s'(N-l)d\right] + i\frac{\Delta\beta_m}{2}\sin\left[s'(N-l)d\right]}, \tag{1}$$

where $s' = \left[(\Delta\beta_m)^2 - \kappa^*\kappa\right]^{\frac{1}{2}}$, $\Delta\beta_m = \beta_m - m\frac{2\pi}{d}$, and $\kappa$ is the coupling coefficient of the two modes. If $\kappa$ is much smaller than $\Delta\beta_m$, Eq. (1) is deduced into

$$E_l = A_0 e^{-i\beta_m(N-l)d}. \tag{2}$$

The interference field of guided modes coming from the $N$ unit cells can be expressed as

$$E'_N = A_0 e^{-i(N-1)\frac{\beta_m d}{2}} \frac{\sin\frac{N\beta_m d}{2}}{\sin\frac{\beta_m d}{2}}. \tag{3}$$

The fields propagate back and forth between the two boundaries, and the multiple reflections give rise to the following superposed field at the right boundary of the $N^{th}$ unit cell

$$E_{b2} = A_0 \frac{\sin\frac{\beta_m N d}{2}}{\sin\frac{\beta_m d}{2}} \frac{r}{1 - r^2 e^{-i\beta_m 2Nd}} e^{-i\frac{(N-1)\beta_m d}{2}} e^{i\varphi_0}. \tag{4}$$

Because the boundary may appear at different positions of the rightmost unit cell, its influence is taken into account by introducing $\varphi_0$. The complex reflection coefficient of the guided waves at either boundary is $r = |r|e^{i\varphi_r}$, where $\varphi_r$ is the phase shift in the reflection. Similarly, the superposed field at the left boundary $E_{b1} = E_{b2} e^{-i2\varphi_0}$. The two waves propagating in opposite directions further interfere with each other in the grating, and the interference field can be written as

$$E = 2A_0 E(x) \frac{\sin \frac{\beta_m N d}{2}}{\sin \frac{\beta_m d}{2}} \frac{r}{1 - r^2 e^{-i\beta_m 2Nd}} e^{-i\beta_m \left(N - \frac{1}{2}\right)d}, \qquad (5)$$

where $E(x) = \cos\left[\beta_m \left(\frac{Nd}{2} - x\right) - \varphi_0\right]$ [see Supplemental Material (SM), S1.1].

When $N \to \infty$, $\beta_m = m \cdot 2\pi/d$. While $N$ is finite, a perturbation on $\beta_m$ is introduced by the boundaries, i.e., the resonant field distributions should satisfy the standing wave condition $\Delta\beta_m Nd - \varphi_r = (m_1 - 1)\pi$, where $m_1$ is an integer. If perfect electric conductors are adopted as the boundaries, $\varphi_r \to \pi$. Therefore, two branches of BICs may be supported by the grating, and $E(x)$ can correspondingly be divided into

$$E_1(x) = \sin\left(\frac{m_1 \pi x}{Nd}\right) \cos\left(\frac{m \cdot 2\pi x}{d}\right) \qquad (6)$$

and

$$E_2(x) = \cos\left(\frac{m_1 \pi x}{Nd}\right) \sin\left(\frac{m \cdot 2\pi x}{d}\right), \qquad (7)$$

respectively [see SM, S1.2].

The intensities of the two BICs can further be expressed as

$$I_1(f) = \frac{|A_0|^2 \sin^2\left(\frac{m_1 \pi x}{Nd}\right) \cos^2\left(\frac{m \cdot 2\pi x}{d}\right) \frac{\sin^2 \frac{\beta_m Nd}{2}}{\sin^2 \frac{\beta_m d}{2}} F}{1 + F \sin^2(\beta_m Nd - \varphi_r)} \qquad (8)$$

and

$$I_2(f) = \frac{|A_0|^2 \cos^2\left(\frac{m_1 \pi x}{Nd}\right) \sin^2\left(\frac{m \cdot 2\pi x}{d}\right) \frac{\sin^2 \frac{\beta_m Nd}{2}}{\sin^2 \frac{\beta_m d}{2}} F}{1 + F \sin^2(\beta_m Nd - \varphi_r)}, \qquad (9)$$

where $R = |r|^2 \to 1$ and $F = 4R/(1-R)^2 \to \infty$. Hence, both branches of the BICs are of infinite intensities at the resonant frequencies and the corresponding Q-factors are also infinite. In practical applications, even if metal mirrors of extremely high reflection are used as the boundaries, the losses caused by wave penetrating the metal are inevitable. Thus, the Q factor becomes finite, which is dependent on $R$, $\varphi_r$, $N$, $m_1$, and $m$ [see SM, S1.3].

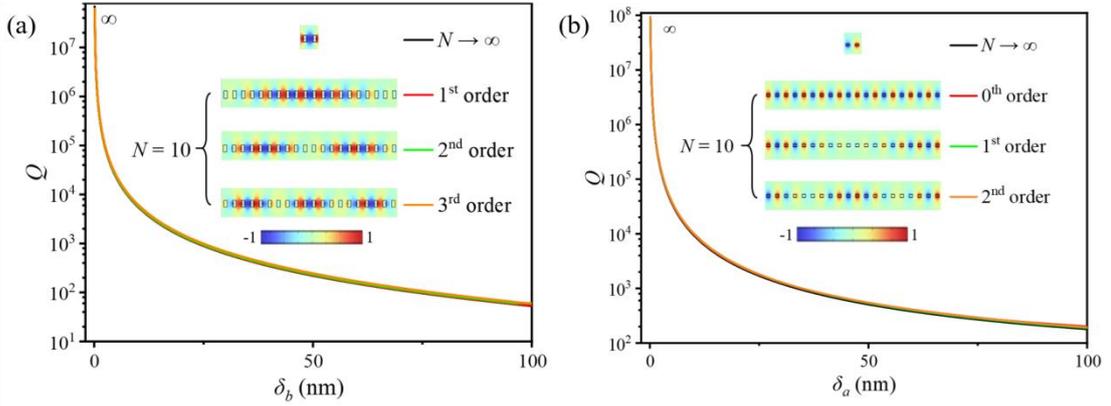

Fig. 2. Dependences of the Q-factor on (a) the geometrical perturbation $\delta_b$ in structures of a height $h_1 = 400$ nm and (b) $\delta_a$ in structures of a height $h_2 = 120$ nm. Both gratings of infinite unit cells and finite unit cells ($N = 10$) with perfect reflection at the boundaries are considered. Insets show the distributions of $y$-polarized electric fields.

To verify the theoretical prediction, the numerical simulations using the finite element method (Comsol Multiphysics) are conducted. Two geometrical structures with the same widths but different heights are employed to obtain two branches of BICs, as shown in Fig. 1. The silicon (Si) is chosen as the dielectric. The widths of Si nanorods and air gaps without perturbations are $a = 200$ nm and $b = 300$ nm, respectively. The widths of air gaps between the outermost nanorods and the boundaries are both set to be $b/2$. The refractive indices of Si and air are 3.42 and 1, respectively. The detailed settings of the simulations are given in SM S2. The eigenfrequency analysis was first performed to calculate the distributions of $y$-polarized electric fields and the Q-factor in the gratings of infinite unit cells ($N \to \infty$, no boundaries) and finite unit cells ($N = 10$) with perfect reflection at the boundaries, respectively. The BIC modes can be divided into two branches due to the introduction of the boundaries. As illustrated in the inset of Fig. 2(a), in structure 1 most power of these BICs is in the air gaps, and the field distributions of the $m_1$th order modes ($m_1 = 1, 2,$ and 3) agree with Eq. (6). The numerical values in the figure are the enhancement factor of the electric field and the integer $m = 1$ in these modes. On the contrary, in structure 2 most power of the BICs is in the Si nanorods, and the field distributions of the $m_1$th order modes ($m_1 = 0$, 1, and 2) are consistent with Eq. (7), as illustrated in the inset of Fig. 2(b). Particularly, in each structure the branch expressed by Eq. (6) or (7) can be supported and can be switched to another by introducing several kinds of geometrical perturbations, such as lateral shift of the cutoff positions of the boundaries by $d/4$, i.e., a nanorod with a width of $a/2$ adjoining either boundary in each structure. Similar to the BICs in the infinitely periodic nanostructure, the Q-factor of the dark BICs in the grating of finite unit cells ($N = 10$) is also infinite, which results from the combination of the symmetry protection and the constructive interference of infinite waves caused by the perfect reflection.

The coupling between waves in free space and the BICs of infinite Q-factor is forbidden due to the symmetry protection. However, by introducing the geometrical perturbations, e.g., shifting one rod in a unit cell away from its original position by $\delta_b$ in structure 1 and increasing the width of one rod in structure 2 by $\delta_a$, as shown in Figs. 1(a) and 1(b), respectively, the symmetry can be broken and the dark BICs transit to bright quasi-BICs of finite Q-factor. The dependences of Q-factor on the two geometrical perturbations are calculated by eigenfrequency analysis, as illustrated in Figs. 2(a) and 2(b). When the geometrical perturbations tend to be zero, the Q-factor approaches to the infinity in the grating of either infinite or finite unit cells, which is the exact result of symmetry-protected BICs. Such symmetry-protected BICs are caused by the destructive interference of diffracted fields from each unit cell to free space. Along with the increase of the geometrical perturbations ($\delta_a$ or $\delta_b$), the Q-factor of quasi-BICs in the gratings of infinite and finite ($N = 10$) unit cells both decreases and their variation curves almost overlap. Thus, infinite unit cells can be mimicked to obtain high Q-factor by introducing the boundaries with perfect reflection in the finite unit cells, which are similar to that of the standing wave grating with amplitude stationary in space but oscillating in time [34].

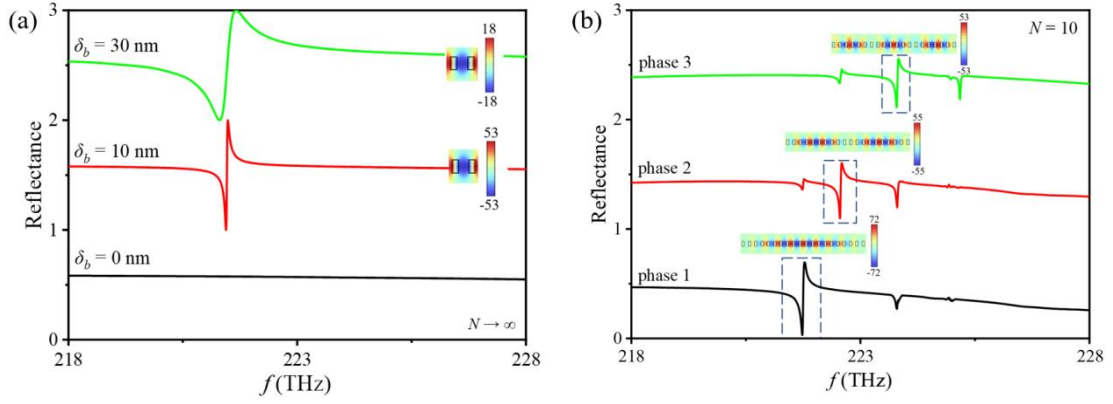

Fig. 3. Reflectance as a function of the frequency under the condition of $h_1 = 400$ nm. (a) Infinite unit cells of different $\delta_b$. (b) Finite unit cells ($N = 10$) of $\delta_b = 10$ nm with perfect reflection at boundaries. Insets show the distributions of $y$-polarized electric fields at the corresponding resonant frequency.

To demonstrate the couple between the waves in free space and the BICs, reflection characteristics of the infinite unit cells and finite unit cells ($N = 10$) with perfect reflection at the boundaries are further simulated. For the infinite grating of a height $h_1 = 400$ nm, uniform plane waves propagating in the negative $z$-direction are adopted as the incident light. As illustrated in Fig. 3(a), when the geometrical perturbation $\delta_b = 0$, no obvious peaks can be observed, corresponding to the symmetry-protected BIC. When the perturbation is introduced, a peak of Fano shape appears due to the symmetry breaking and the peak width increases with $\delta_b$, corresponding to the quasi-BICs. Both the Q-factor calculated from the Fano resonance and the corresponding field distributions are consistent with those obtained by the eigenfrequency analysis shown in Fig. 2(a). The dependence of the peak width on $\delta_b$ for the 10 unit cells is similar to that for the infinite grating. When the perturbation in the 10 unit cells is $\delta_b = 10$ nm, the reflectance spectra are shown in Fig. 3(b). The initial phases of incident waves are adopted as 0, $\sin\left(\dfrac{2\pi x}{Nd}\right)$, and $\sin\left(\dfrac{3\pi x}{Nd}\right)$, which are used to excited the 1st, 2nd, and 3rd quasi-BICs well, respectively. The Q-factor and the field distributions of the Fano resonances are also consistent with those obtained by the eigenfrequency analysis shown in Fig. 2(a). The dependences of Q-factor of the Fano resonances on $\delta_a$ in structures of a height $h_2 = 120$ nm are shown in Fig. S2 in SM, which are similar to those in structures of the height $h_1$ = 400 nm.

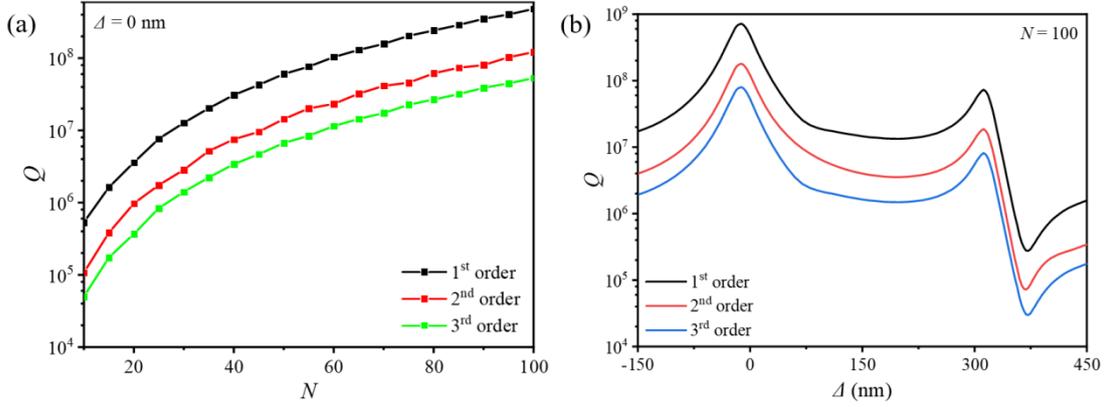

Fig. 4. Dependences of the Q-factor on (a) the number of unit cells and (b) the geometrical perturbation, $\Delta$. The height $h_1$ = 400 nm.

To evaluate the performances of the BICs in actual systems, the boundaries with perfect reflection is then replaced with aluminum (Al) mirrors. The permittivity of Al can be described by the Drude model $\varepsilon_{Al} = 1 - \omega_p^2/\omega(\omega - i\gamma)$, where $\omega$ is angular frequency of light, $\omega_p$ = 2.24×10$^{16}$ rad/s and $\gamma$ = 1.24×10$^{14}$ rad/s are the plasma frequency and damping frequency of Al, respectively [35]. When the perturbation $\delta_b$ = 0 nm and the widths of air gaps between the outermost nanorods and the boundaries $b/2$ = 150 nm in structure 1, the dependences of Q-factor of the $m_1$th order BICs on the number of unit cells, $N$, are calculated. As shown in Fig. 4(a), the Q-factor of the three modes all increases with $N$, which is consistent with the results without considering the influence of the boundaries [27-31]. However, different from them, the Q-factor of the 1$^{st}$ mode can even exceed 10$^8$ when $N$ > 60, which is not proportional to $N^2$ [27-29] or $N^3$ [30-31]. Moreover, the variation curves of the three modes deviate from each other, which mainly results from the difference in field distributions caused by the boundaries.

According to the theory in SM S1.3, the Q-factor of the BICs is also dependent on the phase shift in the boundary reflection, $\varphi_r$. Thus, to adjust the phase equivalently, the geometrical perturbation, $\Delta$, is introduced to the width of air gaps between the outermost nanorods and the boundaries, which is then changed from $b/2$ to $b/2 + \Delta$. The dependences of Q-factor of the three BICs in the 100 unit cells on $\Delta$ is presented in Fig. 4(b). When $\Delta$ varies from −150 to 450 nm, the fluctuating range of the Q-factor can even reach 3 to 4 orders of magnitude. Although the Q-factor is different for the three BICs, their change trends are consistent and the peaks all appear at $\Delta = -13$ nm. As shown in Fig. S3 in SM, the fluctuating range of the Q-factor of the BICs in structure 2 can also reach 3 to 4 orders; however, the peaks arise at $\Delta = 250$ nm.

Although the Q-factor of BICs in the finite grating with metallic mirrors as its boundaries is finite, they remain dark modes due to the symmetry protection. As shown in Figs. S4(a) and S5(a), when the geometrical perturbations $\delta_b$ and $\delta_a$ in structures 1 and 2 tend to be zero, the Q-factor reaches the peaks, which are all finite. However, as shown in Figs. S4(b) and S5(b), without geometrical perturbations the coupling between waves in free space and the BICs of finite Q-factor is still forbidden, which can be excited by introducing the perturbations to break the symmetry.

In summary, a theory based on the superposition principle is established to uncover the basic physics of the wave behavior in a finite grating of $N$ unit cells, which reveals that it can support BICs of infinite Q-factor by introducing perfect reflection at its boundaries. When the boundaries are replaced with metallic mirrors of high reflectivity, the BICs of finite Q-factor are obtained, which can be tuned in a large range by varying the number of unit cells or boundary conditions. In principle the BICs in the two- or three-dimensional array of finite unit cells are of characteristics similar to that in the 1D grating. The results not only provide in-depth understanding of the photonic resonances in finite unit cells, but also have significant application potentials in the design and fabrication of nano-photonic devices.

The authors thank the financial support provided by the National Natural Science Foundation of China under Grant Nos.11874270 and 12174228; Natural Science Foundation of Guangdong Province, China under Grant Nos. 2022A1515011383.

**References:**

1. J. von Neumann and E. Wigner, Über merkwürdige diskrete eigenwerte, Phys. Z 30, 465 (1929).
2. F. H. Stillinger and D. R. Herrick, Bound states in the continuum, Phys. Rev. A 11,

446 (1975).

3. D. V. Evans, M. Levitin and D. Vassiliev, Existence theorems for trapped modes, J. Fluid Mech. 261, 21 (1994).

4. A. A. Lyapina, D. N. Maksimov, A. S. Pilipchuk and A. F. Sadreev, Bound states in the continuum in open acoustic resonators, J. Fluid Mech. 780, 370 (2015).

5. I. Deriy, I. Toftul, M. Petrov, and A. Bogdanov, Bound states in the continuum in compact acoustic resonators, Phys. Rev. Lett. 128, 084301 (2022).

6. M. McIver, An example of non-uniqueness in the two dimensional linear water wave problem, J. Fluid Mech. 315, 257 (1996).

7. M. McIver, Trapped modes supported by submerged obstacles, Proc. R. Soc. A: Math. Phys. Eng. Sci. 456, 1851 (2000).

8. R. Porter, Trapping of water waves by pairs of submerged cylinders, Proc. R. Soc. A: Math. Phys. Eng. Sci. 458, 607 (2002).

9. D. Marinica, A. Borisov, and S. Shabanov, Bound states in the continuum in photonics, Phys. Rev. Lett. 100, 183902 (2008).

10. E. N. Bulgakov and A. F. Sadreev, Bound states in the continuum in photonic waveguides inspired by defects, Phys. Rev. B 78, 075105 (2008).

11. A. Kodigala, T. Lepetit, Q. Gu, B. Bahari, Y. Fainman, and B. Kante, Lasing action from photonic bound states in continuum, Nature 541, 196 (2017).

12. X. Zhang, Y. Liu, J. Han, Y. Kivshar, and Q. Song, Chiral emission from resonant metasurfaces, Science, 377, 1215 (2022).

13. D. Riabov, R. Gladkov, O. Pashina, A. Bogdanov, and S. Makarov, Subwavelength Raman Laser Driven by Quasi Bound State in the Continuum, Laser Photon. Rev. 202300829 (2024).

14. Y. Chen, C. Zhao, Y. Z. Zhang, and C. W. Qiu, Integrated Molar Chiral Sensing Based on High-Q Metasurface, Nano Lett. 20, 8696 (2020).

15. F. Yesilkoy, E. R. Arvelo, Y. Jahani, M. K. Liu, A. Tittl, V. Cevher, Y. Kivshar, and H. Altug, Ultrasensitive hyperspectral imaging and biodetection enabled by dielectric metasurfaces, Nat Photonics 13, 390 (2019).

16. Z. L. Li, G. Z. Nie, Z. Q. Chen, S. P. Zhan, and L. F. Lan, High-quality quasi-bound

state in the continuum enabled single-nanoparticle virus detection, Opt. Lett. 49, 3380 (2024).

17. N. Karl, P. P. Vabishchevich, S. Liu, M. B. Sinclair, G. A. Keeler, G. M. Peake, and I. Brener, All-optical tuning of symmetry protected quasi bound states in the continuum, Appl. Phys. Lett. 115, 141103 (2019).

18. Z. H. Han and Y. J. Cai, All-optical self-switching with ultralow incident laser intensity assisted by a bound state in the continuum, Opt. Lett. 46, 524 (2021).

19. L. Carletti, K. Koshelev, C. De Angelis, and Y. Kivshar, Giant Nonlinear Response at the Nanoscale Driven by Bound States in the Continuum, Phys. Rev. Lett. 121, 033903 (2018).

20. Z. Lin, Y. Xu, Y. Lin, J. Xiang, T. Feng, Q. Cao, J. Li, S. Lan, and J. Liu, High-Q Quasi-bound States in the Continuum for Nonlinear Metasurfaces, Phys. Rev. Lett. 123, 253901 (2019).

21. T. Ning, X. Li, Y. Zhao, L. Yin, Y. Huo, L. Zhao, and Q. Yue, Giant enhancement of harmonic generation in all-dielectric resonant waveguide gratings of quasi-bound states in the continuum, Opt. Express, 28, 34024 (2020).

22. T. Liu, M. Qin, F. Wu, and S. Xiao, High-efficiency optical frequency mixing in an all-dielectric metasurface enabled by multiple bound states in the continuum, Physical Review B, 107, 075441 (2023).

23. M. Parry, A. Mazzanti, A. Poddubny, G. Della Valle, D. Neshev, and A. A. Sukhorukov, "Enhanced generation of non-degenerate photon-pairs in nonlinear metasurfaces", Adv. Phot. 3 (5), 055001 (2021).

24. T. Liu, M. Qin, S. Feng, X. Tu, T. Guo, F. Wu, and S. Xiao, Efficient photon-pair generation empowered by dual quasibound states in the continuum, Phys. Rev. B 109, 155424 (2024)

25. A. Overvig, N. F. Yu, and A. Alù, Chiral Quasi-Bound States in the Continuum, Phys. Rev. Lett. 126 (2021).

26. C. W. Hsu, B. Zhen, A. D. Stone, J. D. Joannopoulos, and M. Soljacic, Bound states in the continuum, Nat. Rev. Mater. 1, 16048 (2016).

27. A. Taghizadeh, and I. S. Chung, Quasi bound states in the continuum with few unit cells of photonic crystal slab, Appl. Phys. Lett. 111, 031114 (2017).

28. E. Bulgakov and A. Sadreev, Propagating Bloch bound states with orbital angular momentum above the light line in the array of dielectric spheres, J. Opt. Soc. Am. A 34, 949 (2017).

29. Z. F. Sadrieva, M. A. Belyakov, M. A. Balezin, P. V. Kapitanova, E. A. Nenasheva, A. F. Sadreev, and A. A. Bogdanov, Experimental observation of a symmetry-protected bound state in the continuum in a chain of dielectric disks, Phys. Rev. A 99, 053804 (2019).

30. E. N. Bulgakov and A. F. Sadreev, High-Q resonant modes in a finite array of dielectric particles, Phys. Rev. A 99, 033851 (2019).

31. M. S. Sidorenko, O. N. Sergaeva, Z. F. Sadrieva, C. Roques-Carmes, P. S. Muraev, D. N. Maksimov, and A. A. Bogdanov, Observation of an Accidental Bound State in the Continuum in a Chain of Dielectric Disks, Phys. Rev. Applied 15, 034041 (2021).

32. E. Bulgakov, G. Shadrina, A. Sadreev, and K. Pichugin, Super-bound states in the continuum through merging in grating, Phys. Rev. B 108, 125303 (2023).

33. A. Yariv and P. Yeh, *Photonics: Optical Electronics in Modern Communications* (6th Edition, Oxford University Press, New York, 2007).

34. S. A. R. Horseley and J. B. Pendry, Travelling Wave Amplification in Stationary Gratings, Phys. Rev. Lett. 133, 156903 (2024).

35. M. A. Ordal, R. J. Bell, R. W. Alexander, L. L. Long, and M. R. Querry, Optical properties of fourteen metals in the infrared and far infrared: Al, Co, Cu, Au, Fe, Pb, Mo, Ni, Pd, Pt, Ag, Ti, V, and W, Appl. Opt. 24, 4493 (1985).

# Supplemental Material

# Bound states in the continuum of infinite quality factor in finite unit cells

Huawei Liang[1], Yuanzhi Liu[1], Yu-Jia Zeng[1], Yangjian Cai[2], Tingyin Ning[2,*]

[1]Key Laboratory of Optoelectronic Devices and Systems of Ministry of Education and Guangdong Province, Shenzhen University, Shenzhen 518060, People's Republic of China

[2]Shandong Provincial Engineering and Technical Center of Light Manipulations & Shandong Provincial Key Laboratory of Optics and Photonic Device, School of Physics and Electronics, Shandong Normal University, Jinan 250358, China

**S1 Theory model**

***S1.1 Generation of* bound states in the continuum (BICs) in finite unit cells with boundaries on both ends**

As shown in Fig. 1, two guided waves with propagation constants of $-\beta_m$ and $\beta_m$ propagate towards the left and right, respectively, where $m$ is the order number of the guided modes. $\kappa$ is the coupling coefficient of the two modes. Their initial amplitudes and phases are assumed $A_0$ and 0 in all unit cells, respectively. According to the coupled-mode theory, when the guided wave from the $l^{th}$ unit cell arrives at the right boundary of the $N^{th}$ unit cell, its electric field can be written as [S1]

$$E_l = A_0 \frac{s'}{s' \cos\left[s'(N-l)d\right] + i\frac{\Delta\beta_m}{2}\sin\left[s'(N-l)d\right]}, \quad (1)$$

where $s' = \left[(\Delta\beta_m)^2 - \kappa^*\kappa\right]^{\frac{1}{2}}$, and $\Delta\beta_m = \beta_m - m\frac{2\pi}{d}$. If $\kappa$ is much smaller than $\Delta\beta_m$, Eq. (1) is deduced into

$$E_l = A_0 e^{-i\beta_m(N-l)d}. \quad (2)$$

At the right boundary of the $N^{th}$ unit cell, the interference field of guided modes coming from the $N$ unit cells can be expressed as

$$E'_N = A_0 e^{-i(N-1)\frac{\beta_m d}{2}} \frac{\sin\frac{N\beta_m d}{2}}{\sin\frac{\beta_m d}{2}}. \tag{3}$$

The fields propagate back and forth between the two boundaries, where multiple reflections happen. Thus, the superposed field at the right boundary of the $N^{th}$ unit cell follows

$$E'_{b2} = A_0 \frac{\sin\frac{\beta_m Nd}{2}}{\sin\frac{\beta_m d}{2}} \frac{r}{1-r^2 e^{-i\beta_m 2Nd}} e^{-i\frac{(N-1)\beta_m d}{2}}, \tag{4}$$

where $r = |r|e^{i\varphi_r}$ is the complex reflection coefficient of the guided waves at the two boundaries, and $\varphi_r$ is the phase shift in the reflection. The right boundary may appear at any position of the rightmost unit cell, and thus Eq. (4) should be changed into

$$E_{b2} = A_0 \frac{\sin\frac{\beta_m Nd}{2}}{\sin\frac{\beta_m d}{2}} \frac{r}{1-r^2 e^{-i\beta_m 2Nd}} e^{-i\frac{(N-1)\beta_m d}{2}} e^{i\varphi_0}, \tag{5}$$

where $\varphi_0$ is related to the influence of the boundary position. Similarly, the superposed field at the left boundary $E_{b1} = E_{b2} e^{-i2\varphi_0}$. The two waves propagating in opposite directions further interfere with each other in the grating, and the fields can be written as

$$E = 2A_0 \cos\left[\beta_m\left(\frac{Nd}{2}-x\right)-\varphi_0\right] \frac{\sin\frac{\beta_m Nd}{2}}{\sin\frac{\beta_m d}{2}} \frac{r}{1-r^2 e^{-i\beta_m 2Nd}} e^{-i\beta_m\left(N-\frac{1}{2}\right)d}. \tag{6}$$

*S1.2 Mode field distributions of the BICs*

When $N \to \infty$, $\beta_m = m \cdot 2\pi/d$. While $N$ is finite, a perturbation on $\beta_m$ is introduced by the boundaries, i.e., $\beta_m = m \cdot 2\pi/d + \Delta\beta_m$. Moreover, the resonant field distributions should satisfy the standing wave condition

$$\Delta\beta_m Nd - \varphi_r = (m_1 - 1)\pi, \tag{7}$$

where $m_1$ is an integer. Thus, Eq. (6) can be changed into

$$E = 2A_0 \cos\left[ mN\pi + \frac{m_1\pi}{2} - \left(\frac{m\cdot 2\pi}{d} + \frac{m_1\pi - \pi + \varphi_r}{Nd}\right)x + \frac{\varphi_r - \pi}{2} - \varphi_0 \right] \times \frac{\sin\frac{\beta_m Nd}{2}}{\sin\frac{\beta_m d}{2}} \frac{r}{1 - r^2 e^{-i\beta_m 2Nd}} e^{-iN\beta_m d} . \quad (8)$$

Let

$$E(x) = \cos\left[ \frac{m_1\pi}{2} - \left(\frac{m\cdot 2\pi}{d} + \frac{m_1\pi - \pi + \varphi_r}{Nd}\right)x + \frac{\varphi_r - \pi}{2} - \varphi_0 \right], \quad (9)$$

which is used to discuss the dependence of the mode fields on $x$. Eq. (9) can be written as

$$E(x) = \cos\left( \frac{m_1\pi}{2} - \frac{m_1\pi - \pi + \varphi_r}{Nd} x \right) \cos\left( \frac{m\cdot 2\pi}{d} x - \frac{\varphi_r - \pi}{2} + \varphi_0 \right) \\ + \sin\left( \frac{m_1\pi}{2} - \frac{m_1\pi - \pi + \varphi_r}{Nd} x \right) \sin\left( \frac{m\cdot 2\pi}{d} x - \frac{\varphi_r - \pi}{2} + \varphi_0 \right) . \quad (10)$$

If perfect electrical conductors are applied to the boundaries, $\varphi_r \to \pi$ and the tangential component of the electric field at the boundaries is zero. Therefore, when $\sin\left(\frac{m\cdot 2\pi x}{d} + \varphi_0\right) = 0$ at the positions of $x = 0$ and $x = Nd$, $\cos\left(\frac{m_1\pi}{2} - \frac{m_1\pi x}{Nd}\right) = 0$.

Similarly, when $\cos\left(\frac{m\cdot 2\pi x}{d} + \varphi_0\right) = 0$, $\sin\left(\frac{m_1\pi}{2} - \frac{m_1\pi x}{Nd}\right) = 0$ at the two positions.

As a result, Eq. (10) can be divided into two parts, i.e.,

$$E_1(x) = \pm \sin\left(\frac{m_1\pi x}{Nd}\right) \cos\left(\frac{m\cdot 2\pi x}{d}\right), \quad (11)$$

and

$$E_2(x) = \pm \cos\left(\frac{m_1\pi x}{Nd}\right) \sin\left(\frac{m\cdot 2\pi x}{d}\right). \quad (12)$$

The electric fields of the corresponding two BICs can further be written as

$$E_1 = 2A_0 \sin\left(\frac{m_1\pi x}{Nd}\right) \cos\left(\frac{m\cdot 2\pi x}{d}\right) \frac{\sin\frac{\beta_m Nd}{2}}{\sin\frac{\beta_m d}{2}} \frac{r}{1 - r^2 e^{-i\beta_m 2Nd}} e^{-iN\beta_m d}, \quad (13)$$

and

$$E_2 = 2A_0 \cos\left(\frac{m_1 \pi x}{Nd}\right) \sin\left(\frac{m \cdot 2\pi x}{d}\right) \frac{\sin\frac{\beta_m Nd}{2}}{\sin\frac{\beta_m d}{2}} \frac{r}{1-r^2 e^{-i\beta_m 2Nd}} e^{-iN\beta_m d}, \quad (14)$$

respectively.

### S1.3 Q-factor of the BICs

The intensities of the two BICs can be expressed as

$$I_1(f) = \frac{|A_0|^2 \sin^2\left(\frac{m_1 \pi x}{Nd}\right) \cos^2\left(\frac{m \cdot 2\pi x}{d}\right) \frac{\sin^2 \frac{\beta_m Nd}{2}}{\sin^2 \frac{\beta_m d}{2}} F}{1 + F \sin^2(\beta_m Nd - \varphi_r)}, \quad (15)$$

and

$$I_2(f) = \frac{|A_0|^2 \cos^2\left(\frac{m_1 \pi x}{Nd}\right) \sin^2\left(\frac{m \cdot 2\pi x}{d}\right) \frac{\sin^2 \frac{\beta_m Nd}{2}}{\sin^2 \frac{\beta_m d}{2}} F}{1 + F \sin^2(\beta_m Nd - \varphi_r)}, \quad (16)$$

where $R = |r|^2$ and $F = 4R/(1-R)^2$. Perfect electric conductors are adopted as the boundaries, so $R \to 1$ and $F \to \infty$. According to Eqs. (15) and (16), both branches of the BICs are of infinite intensities at the resonant frequencies and the corresponding quality Q-factors are also infinite. Thus, infinite unit cells with infinite Q-factor can be mimicked by introducing the boundaries with perfect reflection in the finite unit cells

When metal mirrors are used as the boundaries, the losses caused by wave penetrating the metal are inevitable, and the Q factor correspondingly becomes finite. At the frequency near the resonant points Eqs. (15) and (16) can be changed to

$$I_1(f) = \frac{|A_0|^2 E_1^2(x) \frac{\sin^2 \frac{\beta_m Nd + \delta\beta_m Nd}{2}}{\sin^2 \frac{\beta_m Nd + \delta\beta_m Nd}{2N}} F}{1 + F \sin^2(\delta\beta_m Nd)}, \quad (17)$$

and

$$I_2(f) = \frac{|A_0|^2 E_2^2(x) \dfrac{\sin^2 \dfrac{\beta_m Nd + \delta\beta_m Nd}{2}}{\sin^2 \dfrac{\beta_m Nd + \delta\beta_m Nd}{2}} F}{1 + F \sin^2 (\delta\beta_m Nd)}, \tag{18}$$

where $\delta\beta_m = \beta_m(f) - \beta_m$, and $f$ is the frequency. $\beta_m(f)$ is approximately proportional to the wave number, $k$, where $k = \dfrac{2\pi f}{c}$, and $c$ is the velocity of light in vacuum. Therefore, the Q-factor can be expressed as

$$Q = \frac{f_m}{\delta f_{\text{FWHM}}} = \frac{\beta_m}{\delta\beta_{m,\text{FWHM}}}, \tag{19}$$

where $f_m$ is the resonant frequency and $\delta f_{\text{FWHM}}$ is the full width at half maximum of the frequency. $\delta\beta_{m,\text{FWHM}}$ is the corresponding width of the propagation constant. According to Eqs. (17), (18) and (19), the Q-factor of both BICs is decided by

$$\frac{2\sin^2 \dfrac{m_1\pi - \pi + \varphi_r}{2N} \sin^2 \left[\dfrac{Nm2\pi + m_1\pi - \pi + \varphi_r}{2}\left(1 + \dfrac{1}{2Q_1}\right)\right]}{\sin^2 \dfrac{m_1\pi - \pi + \varphi_r}{2} \sin^2 \left[\dfrac{Nm2\pi + m_1\pi - \pi + \varphi_r}{2N}\left(1 + \dfrac{1}{2Q_1}\right)\right]} = 1 + F \sin^2 \dfrac{Nm2\pi + m_1\pi - \pi + \varphi_r}{2Q_1}. \tag{20}$$

As illustrated in Eq. (20), the Q-factor is dependent on $R$, $\varphi_r$, $N$, $m_1$, and $m$. Thus, in addition to the increase of the reflectivity at the boundaries and the number of the unit cells, the Q-factor can be improved by adjusting the phase delay in the boundary reflection, $\varphi_r$, which may be realized equivalently by introducing the geometrical perturbation to the width of air gaps between the outermost nanorods and the boundaries.

**S2 Simulation settings**

Different simulation models using the Comsol Multiphysics are schematically shown in Fig. S1. Figure S1(a) presents the model of a unit cell, where periodic boundary conditions (PBCs) are employed on both sides to mimic the grating of infinite unit cells. The perfect matched layers (PMLs) are used on the top and bottom domains to reduce the influence of wave reflection. The eigenfrequency analysis is first performed to obtain the resonance mode and Q-factor. Subsequently a plane-wave radiating from Port 1 is incident normally upon the grating, and the reflected and transmitted waves are absorbed by Ports 1 and 2, respectively. The S-parameter of Port 1, $S_{11}$, can be used to calculate the reflection spectra by sweeping the frequency. For the grating of finite unit cells, the PBCs are replaced with the perfect electric conductors (PECs) and aluminum (Al) layers, respectively, as shown in Fig. S1(b) and S1(c), but the PML layers are still employed on the top and bottom domains. The eigenfrequency analysis and the method to obtain reflection spectra are similar to that for infinite unit cells.

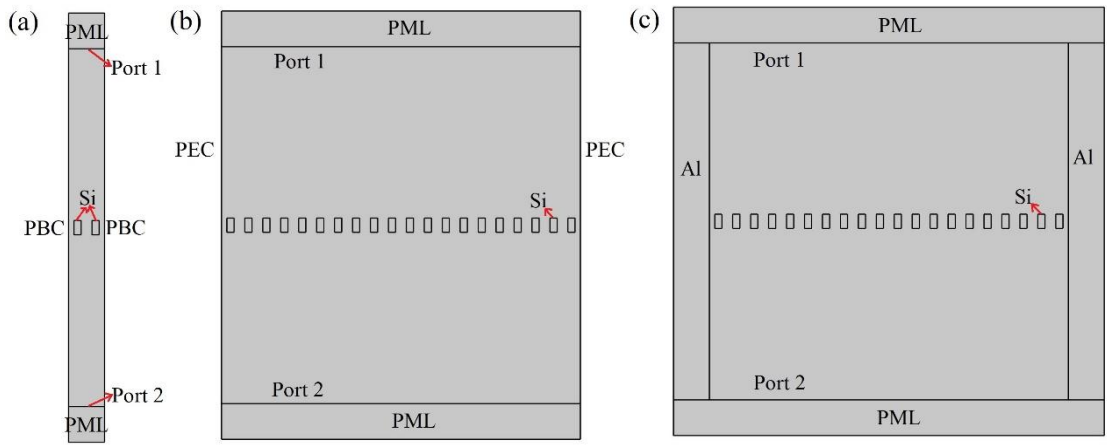

Fig. S1. Schematics of different simulation models. (a) Infinite unit cells, (b) finite unit cells ($N = 10$) with PEC at the two boundaries, and (c) finite unit cells ($N = 10$) with Al layers at the two boundaries.

**S3 Simulated Results of structure 2**

In structure 2, the widths of one Si nanorod and one air gap remain unchanged, while the widths of the other Si nanorod and air gap become $a + \delta_a$, and $b - \delta_a$ by introducing a perturbation $\delta_a$, as shown in Fig. 1. The height of Si nanorods is $h_2$=120 nm.

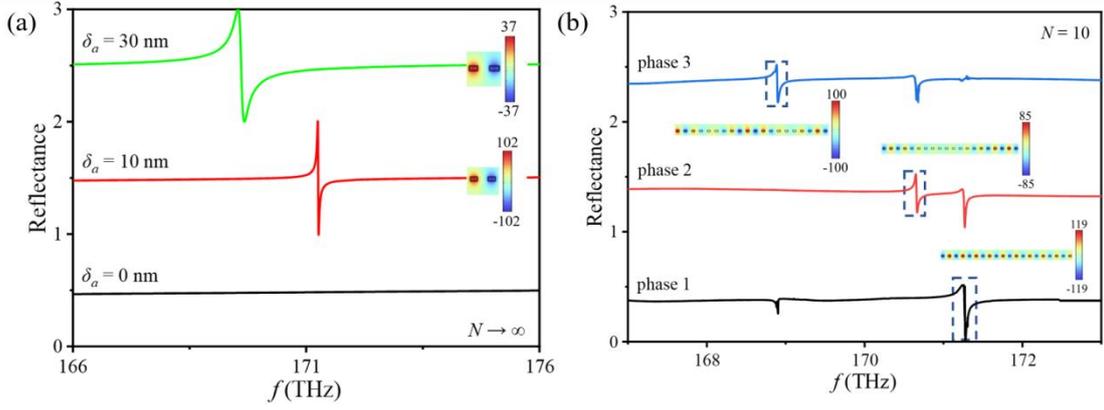

Fig. S2. Dependences of the reflectance on the frequency. (a) Infinite unit cells of different $\delta_a$. The incident wave is a uniform plane wave. (b) Finite unit cells ($N$=10) of $\delta_a$ =10 nm with perfect reflection at boundaries. Phases 1, 2, and 3 of incident waves are 0, $\sin\left(\dfrac{2\pi x}{Nd}\right)$, and $\sin\left(\dfrac{3\pi x}{Nd}\right)$, respectively. The insets show the distributions of $y$-polarized electric fields at the corresponding resonant frequency.

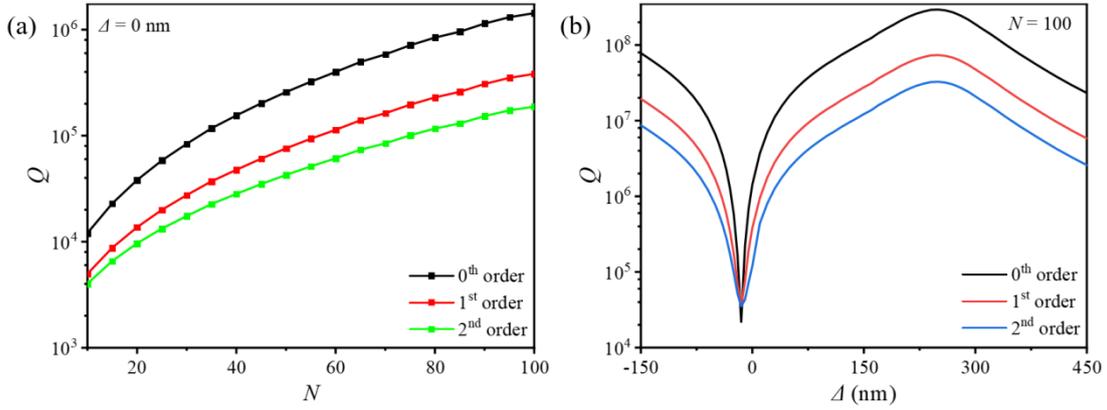

Fig. S3. Q-factor as a function of (a) the number of unit cells and (b) the geometrical perturbation, $\varDelta$.

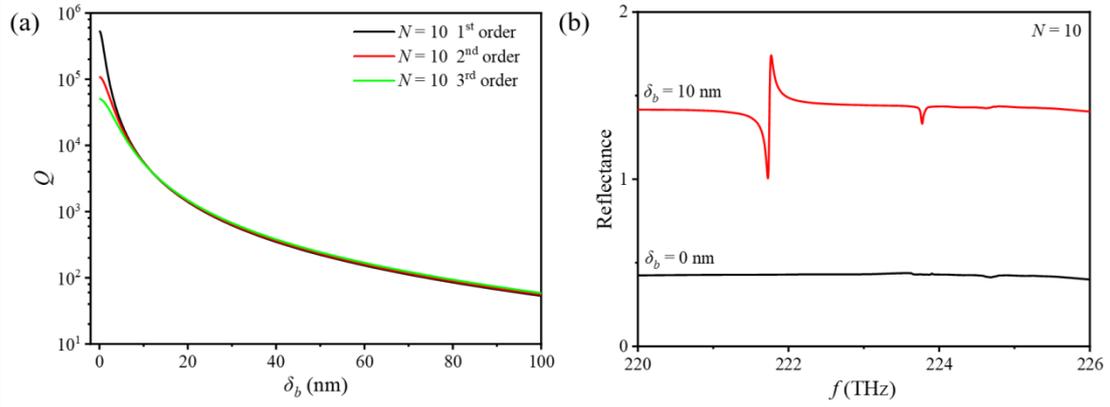

Fig. S4. Dependences of (a) the Q-factor on the geometrical perturbation, $\delta_b$, and (b) the reflectance on the frequency in structure 1. The boundary materials are Al.

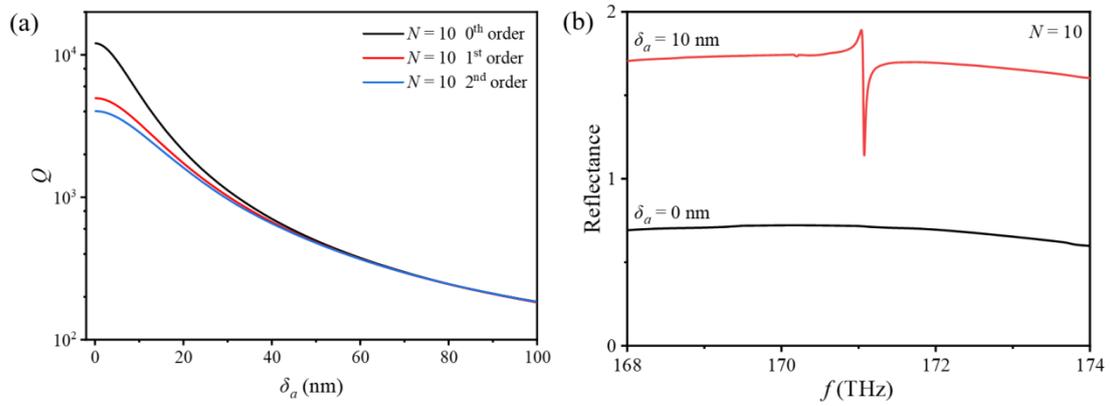

Fig. S5. Dependences of (a) the Q-factor on the geometrical perturbation, $\delta_a$, and (b) the reflectance on the frequency in structure 2. The boundary materials are Al.

[S1] A. Yariv and P. Yeh, *Photonics: Optical Electronics in Modern Communications* (6th Edition, Oxford University Press, New York, 2007).